\documentclass[]{paper}
\usepackage{color}
\usepackage{amsmath, graphics,
mathrsfs,latexsym,amscd,graphicx,amssymb, bm, upgreek}

\newcommand{\be}{\begin{equation}}
\newcommand{\ee}{\end{equation}}
\newcommand{\ba}{\begin{eqnarray}}
\newcommand{\ea}{\end{eqnarray}}

\newcommand{\pa}{\partial}
\newcommand{\f}{\frac}

\title{ Variational derivation of two-component Camassa-Holm shallow water system
}

\author{\normalsize Delia IONESCU-KRUSE\\
\normalsize Institute of Mathematics of
the Romanian Academy,\\
\normalsize P.O. Box 1-764, RO-014700, Bucharest,
 Romania\\
\normalsize E-mail: Delia.Ionescu@imar.ro\\[10pt]}

 \date{}

\begin{document}
\maketitle

\begin{abstract}
By a variational approach in the Lagrangian formalism, we derive
the nonlinear integrable two-component Camassa-Holm system
(\ref{2ch}). We show that the two-component Camassa-Holm system
(\ref{2ch}) with the plus sign arises as an approximation to the
Euler equations of hydrodynamics for propagation of irrotational
shallow water waves over a flat bed. The Lagrangian used in the
variational derivation is not a metric.
\bigskip

\begin{keywords}
two-component Camassa-Holm system, shallow water waves,
variational methods.\\
\textbf{AMS Subject Classification}: 35Q53, 76B15, 70G75 
\end{keywords}

\end{abstract}

\section{Introduction}

A two-component generalization of the peakon Camassa-Holm equation
 is given by
\begin{equation}
\left\{\begin{array}{ll}
u_t+3uu_x-u_{txx}-2u_xu_{xx}-uu_{xxx}\pm HH_x=0\\
\\
 H_t+(Hu)_x=0
\end{array}
\right. \label{2ch}
\end{equation} with $x\in\mathbf{R}$, $t\in\mathbf{R}$,
$u(x,t)\in \mathbf{R}$, $H(x,t)\in \mathbf{R}$. Subscripts here,
and later, denote partial derivatives. Obviously, for $H=0$, the
system (\ref{2ch}) reduces to  the peakon equation derived by
Camassa and Holm \cite{ch} (for alternative derivations within the
shallow water regime see also Refs. \cite{johnson1},
\cite{dullin}, \cite{io}, \cite{c&l}).

 The choice of the minus sign
  is often considered in (\ref{2ch}). This system appears
  originally in
   \cite{olver&ros},
where  it is shown that most integrable bi-Hamiltonian systems are
governed
   by tri-Hamiltonian structures. Then, by recombining the Hamiltonian operators
one is lead to integrable hierarchies,  one of this integrable
hierarchy including the system (\ref{2ch}) with minus sign.
Alternative derivations of the system (\ref{2ch}) with the minus
sign are provided in \cite{shabat}, \cite{liu&zhang},
\cite{chen&liu&zhang}, \cite{falqui}. In \cite{chen&liu&zhang} is
established an explicit reciprocal transformation between
(\ref{2ch}), called here the 2-CH system, and the first negative
flow of the AKNS hierarchy. Thus,  solutions of the 2-CH system
are obtained from solutions of the first negative flow of the AKNS
hierarchy, among them the peakon and multi-kink solutions. In
\cite{falqui} it is also given a Lax pair formulation of the
system (\ref{2ch}). So far, there is no physical interpretation of
the system (\ref{2ch}) with the choice of the minus sign.

For the choice of the plus sign in (\ref{2ch}), the system may be
regarded as a model of shallow water waves. In \cite{const&iv}
this system is derived from the Green-Naghdi
equations\footnote{ For more
background information on the Green-Naghdi equations, see, for
example, the discussion in the recent book \cite{const-carte}.}
\cite{green&naghdi} by  using expansions of the variable in terms of
physical parameters. There it is also noticed that, with the
asymptotic limits $u\rightarrow 0$ and $H\rightarrow 1$ as
$|x|\rightarrow \infty$, the peakon solution is absent among the
solitary waves solutions of (\ref{2ch}) with the plus sign. A
modified two-component Camassa-Holm  system is proposed in
\cite{honatr}. This modified system, derived in \cite{honatr} as
semidirect-product Euler-Poincar\' e equations \cite{homara} with
the Lagrangian given by a $H^1$ metric in $u$ and $H^1$ metric in
$(H-1)$, allows singular solutions, but may not be integrable.

The mathematical properties of the system (\ref{2ch}) have been
studied further in many works. Its analytical properties such as
well-posedness and wave breaking were studied in \cite{escher},
\cite{const&iv}, \cite{guan&yin} and others; more
about its geometric properties see in \cite{ekl}. For
investigations of the traveling wave solutions  we refer to
\cite{const&iv}, \cite{mustafa}, \cite{mohajer}.

The purpose of the present paper is to derive by a variational
approach in the Lagrangian formalism, the two-component
Camassa-Holm shallow water system, that is, the system (\ref{2ch})
with the plus sign. Starting from a general dimensionless version
of the irrotational water wave equations, we introduce the two
fundamental parameters $\epsilon$ and $\delta$, associated with
amplitude and with wave length. The shallow water approximation is
obtained by requiring $\delta\rightarrow 0$, for arbitrary fixed
$\epsilon$. In this regime, the leading-order problem reduces to a
system of two equations on the surface elevation and the
horizontal component of the velocity, that is, the system
(\ref{sw1}), or (\ref{sw2}) in view of the notation
(\ref{notation}). These equations are the so-called classical
shallow water equations (see, for example, \cite{stoker}).  We are
looking for a higher-order correction to the classical shallow
water equations. The second equation in (\ref{sw2}), which can be
interpreted as a continuity equation for the free surface
elevation, is exactly the second equation of the system
(\ref{2ch}). The first equation of the system (\ref{2ch}) with
plus sign, will be an Euler-Lagrange equation yielding the
critical points of an action (functional) in the space of path,
within the Lagrangian formalism. Firstly, we approximate the
kinetic energy at the free surface. We consider the parameter
$\epsilon$ such that $\epsilon^3$ and $\epsilon^4$ contributions
in the kinetic energy at the free surface (\ref{5}) can be
neglected.  In the Eulerian picture, the Lagrangian is defined as
the approximated kinetic energy at the free surface (\ref{energy})
minus the potential energy calculated with respect to the
undisturbed water level (\ref{21}). In the Lagrangian formalism,
the Lagrangian will be obtained from the Lagrangian in the
Eulerian picture, taking also into account the second equation of
the system (\ref{2ch}), see (\ref{lagr}). This Lagrangian is not a
metric. To the order of our approximation, we get that the
non-dimensional horizontal velocity
 of the water $u$ and the non-dimensional free upper surface
 $H$,
  satisfy the Camassa-Holm shallow water system (\ref{2ch}) with the plus sign,
  see Theorem 1.
   We also observe that, the Camassa-Holm  system (\ref{2ch}) with the
   minus sign
   is obtained if instead of the Lagrangian (\ref{lagreuler}) we
   consider the total energy at the free surface,
  that is,  the approximated kinetic energy at the free surface (\ref{energy})
plus the potential energy calculated with respect to the
undisturbed water level (\ref{21}). This Lagrangian is a metric,
namely, a $H^1$ metric in $u$ and $L^2$ metric in $(H-1)$. We
point out that (\ref{2ch}) with the minus sign can be regarded as
geodesic equations on the semidirect product of Diff$(\mathbf{R})$
with $\mathcal{F}(\mathbf{R}$) (see also \cite{honatr}).

The derivation of the Camassa-Holm shallow water
system (\ref{2ch}) with the plus sign is made under the assumption
of irrotational flow. There are circumstances in which this
assumption is well justified but there are cases where it is
inappropriate. Tidal flow is a well-known example when constant
vorticity flow is an appropriate model (see \cite{dasilva}; for a
discussion of the physical relevance of flows with constant
vorticity see also \cite{c2011}). The derivation of a
two-component system modelling shallow water waves  in the
constant vorticity case, by introducing of suitable scalings and
by truncating asymptotic expansions of the quantities to
appropriate order,  is given in \cite{rosen}. The derivation of
such a model by a variational approach in the Lagrangian formalism
will be presented in a future paper.


\section{Nondimensionalisation of the governing equations for water waves.
 Classical shallow water equations}
 The two-dimensional gravity water
waves moving over
 an irrotational flow are described by the following boundary value problem:
\begin{equation}
\begin{array}{c}
\begin{array}{ll}
u_t+uu_x+vu_z=- p_x\\ \,\, v_t+uv_x+vv_z=- p_z-g\\
\end{array}
\quad \quad \quad \quad \textrm{ (EEs) }\\
\\

 \qquad \qquad u_x+v_z=0  \qquad \qquad \qquad \qquad \textrm{ (MC)  }\\
\\

 \qquad \qquad u_z-v_x=0  \qquad \qquad \qquad \quad \textrm{ (IC)  }\\
\\

\begin{array}{ll}
  v=\eta_t+u\eta_x \, \, \textrm{ on }\,
z=h_0+\eta(x,t)\\
\qquad \quad v=0 \, \, \textrm { on } z=0
\end{array}
\quad \,\,\, \textrm{ (KBCs) }\\
\\

\qquad
 p=p_0 \,  \textrm{ on } z=h_0+\eta(x,t)
  \quad \quad \textrm{ (DBC)} \end{array} \label{e+bc}
\ee  where $(u(x,z,t), v(x,z,t))$ is the velocity field of the
water - no motion takes place in the $y$-direction, $p(x,z,t)$
denotes the pressure, $g$ is the constant gravitational
acceleration, $p_0$ being  the constant atmospheric pressure. The
water moves in a domain with a free upper surface at
$z=h_0+\eta(x,t)$, for a constant $h_0>0$,  and a flat bottom at
$z=0$. The undisturbed water surface is $z=h_0$ and $\eta(x,t)$ is
the displacement of the free surface from the undisturbed state.
Euler's equations (EEs) are the appropriate equations of motion
for gravity water waves, see \cite{johnson-carte}. Another
realistic assumption for gravity water wave problem is the
incompressibility, that is, the density $\rho$ is constant (see
\cite{lighthill}), which implies the equation of mass conservation
(MC). We set the constant water density $\rho=1$.  The
idealization of irrotational flow is physically relevant in the
absence of non-uniform currents in the water. The zero vorticity
flow is characterized by the additional equation (IC). The
kinematic boundary conditions (KBCs) express the fact that the
same particles always form the free-water surface and that the
fluid is assumed to be bounded below by a hard horizontal bed
$z=0$. The dynamic boundary condition (DBC) expresses the fact
that on the free surface the pressure is equal to the constant
atmospheric pressure denoted $p_0$.\\
\\

\hspace{1.5cm}\scalebox{0.65}{\includegraphics{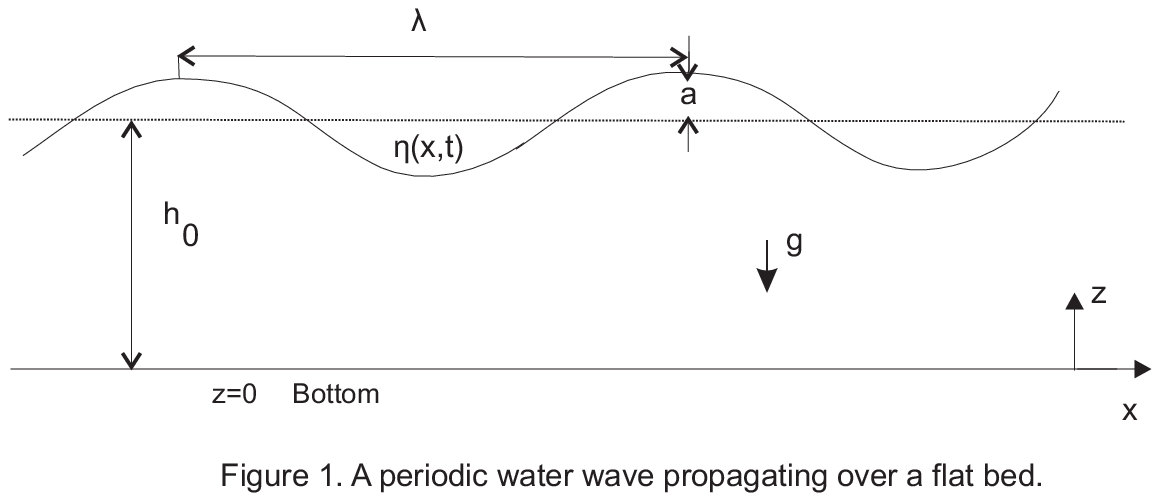}}\\
\\

 We
non-dimensionalize the set of equations (\ref{e+bc}) using the
undisturbed depth of the water $h_0$, as the vertical scale, a
typical wavelength $\lambda$, as the horizontal scale, and a
typical amplitude of the surface wave $a$ (see Figure 1). Thus, we
define the set of non-dimensional variables (for more details see
\cite{johnson-carte}):
\begin{equation}
\begin{array}{c}
x\mapsto\lambda x,  \quad z\mapsto h_0 z, \quad \eta\mapsto a\eta,
\quad t\mapsto\f\lambda{\sqrt{gh_0}}t,\\
 \\

  u\mapsto  \sqrt{gh_0}u,
\quad v\mapsto h_0\f{\sqrt{gh_0}}{\lambda}v,\\
 \\

 p\mapsto p_0+ g h_0(1-z)+ g h_0 p,
 \end{array} \label{nondim}\end{equation}
where, to avoid new notations, we have used the same symbols for
the non-dimensional variables  $x$, $z$, $\eta$, $t$, $u$, $v$,
$p$ on the right-hand side. Therefore, in the non-dimensional
variables (\ref{nondim}), the water-wave problem (\ref{e+bc})
becomes:
\begin{equation}
\begin{array}{cc}
u_t+uu_x+vu_z=- p_x&\\  \delta^2(v_t+uv_x+vv_z)=- p_z&\\
 u_x+v_z=0&\\
 u_z-\delta^2v_x=0&\\
v=\epsilon(\eta_t+u\eta_x) & \textrm{ on }\,
z=1+\epsilon\eta(x,t)\\
 p=\epsilon\eta&
\textrm{ on }\,
z=1+\epsilon\eta(x,t)\\
 v=0 &
\textrm { on } \, z=0
 \end{array}
\label{e+bc'} \end{equation}  where we have introduced the
amplitude parameter $\epsilon=\f a{h_0}$ and the shallowness
parameter $\delta=\f {h_0}{\lambda}$.

The shallow water approximation is  obtained by requiring
$\delta\rightarrow 0$, for arbitrary fixed $\epsilon$. For
$\delta=0$, the leading-order equations become:
\begin{equation}
\begin{array}{cc}
u_t+uu_x+vu_z=- p_x&\\
p_z=0&\\
 u_x+v_z=0&\\
 u_z=0&\\
v=\epsilon(\eta_t+u\eta_x) \,& \textrm{ on }\,
z=1+\epsilon\eta(x,t) \\
p=\epsilon\eta(x,t)\,&  \textrm{ on }\, z=1+\epsilon\eta(x,t)\\
v=0 \,& \textrm { on } z=0
\end{array}
\label{small}
\end{equation}
From the second equation and the forth equation in (\ref{small}),
we get that $u$ and $p$  do not depend on $z$. Thus, \be u=u(x,t),
\label{13}\ee and because $p=\epsilon\eta(x,t)$ on
$z=1+\epsilon\eta(x,t)$, we get
\begin{equation} p=\epsilon\eta(x,t). \label{2}\end{equation} Taking into
account (\ref{13})  and  the third equation in (\ref{small})  we
obtain \be v=-zu_x+g(x,t), \ee $g(x,t)$ being an arbitrary
function. Further, by the last equation in (\ref{small}), we have
\be v=-zu_x. \label{3} \ee With (\ref{13}) and  (\ref{2}) in view,
the first equation  in (\ref{small}) can be written as \be u_t+ u
u_x+\epsilon\eta_x=0. \label{4}\ee From (\ref{3}), the fifth
equation in (\ref{small}) becomes \be \epsilon\eta_t+\epsilon
u\eta_x+u_x+\epsilon\eta u_x=0. \ee Hence, for $\delta=0$, the
system which describes the evolution of the surface waves  is
\be\left\{
\begin{array}{ll}
u_t+ u u_x+\epsilon\eta_x=0\\
\epsilon \eta_t+[(1+\epsilon\eta)u]_x=0,
\end{array}
\right.\label{sw1} \ee the other two variables $p$ and $v$ are
expressed function of $\eta$ and $u$, see (\ref{2}), respectively
(\ref{3}). In the study of the full governing
equations for water waves, the relations between the free surface,
the fluid velocity and the pressure are more subtle. Various
properties of the pressure and the velocity beneath Stokes waves
and deep-water Stokes waves are proved, by methods relying upon
harmonic and superharmonic functions,
in \cite{CS}, respectively \cite{henry}.  \\
 If we
denote by \be H(x,t):=1+\epsilon\eta(x,t)\label{notation} \ee
then, the system of equations (\ref{sw1}) becomes \be\left\{
\begin{array}{ll}
u_t+ u u_x+H_x=0\\
H_t+(H u)_x=0. \end{array} \right.\label{sw2} \ee The set of
hyperbolic partial differential equations (\ref{sw2})
are the so-called classical shallow water equations (see, for
example, \cite{stoker}).
The equations (\ref{sw2}) can be written in Hamiltonian
form relative to a symplectic structure introduced by Manin
\cite{manin}.
The second Hamiltonian structure for the system (\ref{sw2}) was
obtained by  Cavalcante and McKean \cite{caval&mckean}. In fact,
the system (\ref{sw2}), as a particular case of the system of
polytropic gas equations in 1 + 1, i.e. in one spatial and one
temporal dimension, is Hamiltonian with respect to three distinct
Hamiltonian structures \cite{nutku}. These Hamiltonian structures
are compatible and thus, the system of equations (\ref{sw2}) is
completely integrable \cite{olver}. The infinite sequence of
integrals of motion (the conserved quantities) that the
Hamiltonian structures give rise to was found much earlier by
Benney \cite{benney}.  
The
equations (\ref{sw2}) provide a good approximation to the exact
solution of the water-wave problem; for a rigorous  justification
with a precise control of the estimated error see \cite{lannes1}.

\section{Approximation procedure and  variational derivation of the two-component
 Camassa-Holm shallow water system}

 By an interplay of small-parameter expansions and variational methods
 we derive  the two-component
 Camassa-Holm shallow water equations.

 The two important  parameters $\epsilon$ and $\delta$  that arise
in water-waves theories in the forms $\epsilon$ and $\delta^2$,
are used to define various approximations of the governing
equations and the boundary conditions. The role of $\delta$
independent of $\epsilon$ is useful in the description of
arbitrary amplitude shallow water waves, that is,
$\delta\rightarrow 0$, $\epsilon$ fixed. The small-amplitude
shallow water  waves are obtained in the limits
$\epsilon\rightarrow 0$ and $\delta\rightarrow 0$.

 We observe  in (\ref{e+bc'}) that on
$z=1+\epsilon\eta$ both $v$ and $p$ are proportional to
$\epsilon$, this being consistent with the fact that as
$\epsilon\rightarrow 0$ we must have $v\rightarrow 0$ and
$p\rightarrow 0$ (with no disturbance the free
surface becomes a
horizontal surface on which $v=p=0$).  For consistence, one
requires that $u$ is proportional to $\epsilon$ too. $\epsilon$ is
also called the nonlinearity parameter. Indeed, we see that the
terms $uu_x$ and $\epsilon \eta u$ from (\ref{sw1}), are
proportional with $\epsilon^2$ and by neglecting these terms, the
system (\ref{sw1}) reduces to the linear system\be\left\{
\begin{array}{ll}
u_t+\epsilon\eta_x=0\\
\epsilon \eta_t+u_x=0.
\end{array}
\right.\label{sw1'} \ee Therefore,  $\eta(x,t)$  satisfies the
linear wave equation \be \eta_{tt}-\eta_{xx}=0. \label{linear}\ee
By choosing the right-going wave of (\ref{linear}), that is, \be
\eta(x,t)=f(x-t), \ee we obtain the solution \be
u=\epsilon\eta(x,t)+\textrm{const},\, v=-\epsilon z\eta_x(x,t),\,
p=\epsilon \eta(x,t). \ee

 We are looking for a higher-order correction to the
  classical shallow water equations (\ref{sw1}),
or (\ref{sw2}) in view of the notation (\ref{notation}). We
observe that the second equation in (\ref{sw2}) is exactly the
second equation of the two-component
 Camassa-Holm shallow water system (\ref{2ch}).  In order to get the two-component
 Camassa-Holm shallow water system (\ref{2ch}) with the plus sign,
  instead of making asymptotic expansions
 in the equations of motion, we  approximate the kinetic energy  at the free surface
 and we use the variational methods
 in the Lagrangian formalism.

 Taking into account the
components (\ref{13}) and (\ref{3}) of the velocity field,  the
kinetic energy has at the free surface $z=1+\epsilon\eta(x,t)$ the
expression
 \be E_c(u,\eta)=\frac 1{2}
\int_{-\infty}^\infty[u^2+(1+\epsilon\eta)^2u^2_x]dx. \label{5}\ee
As we saw above, the non-dimensional $u$ is proportional to
$\epsilon$. We consider the parameter $\epsilon$ such that
$\epsilon^3$ and $\epsilon^4$ contributions in the right hand side
of (\ref{5}) can be neglected. Thus, we approximate the kinetic
energy at the free surface by \ba E_c&\approx &\frac 1{2}
\int_{-\infty}^\infty[u^2+u^2_x]dx=:E_c(u). \label{energy}\ea
 We require in (\ref{energy})
that $u(x,t)$ and $u_x(x,t)$  decay rapidly at $\pm$ infinity, at
any instant $t$.\\
In non-dimensional variables, with $\rho$ and $g$ settled at 1, we
define the gravitational potential energy at the free surface
$z=1+\epsilon\eta(x,t)$, gained by the fluid parcel when it is
vertically displaced from its undisturbed position with $\epsilon
\eta(x,t)$, by \ba E_p(\eta)&=&
\int_{-\infty}^\infty\left(\int_0^{1+\epsilon\eta}(z-1)\,dz\right)dx=\frac
1{2} \int_{-\infty}^\infty(\epsilon\eta)^2dx\nonumber\\
&\stackrel{(\ref{notation})}{=}& \f1{2}\int_{-\infty}^\infty
(H-1)^2dx=:E_p(H). \label{21}\ea We require in  (\ref{21}) that
$H(x,t)\rightarrow 1$  as $x\rightarrow\pm \infty$, at any instant
$t$.

 The
second equation in (\ref{sw2}) was derived by replacing in  the
free-surface kinematic boundary condition, that is, the fifth
equation in (\ref{small}), the vertical velocity $v$ obtained from
the mass conservation equation, that is, the third equation in
(\ref{small}), for an horizontal velocity $u$ given by (\ref{13})
and taking also into account that the rigid bottom is
impenetrable.  The Lagrangian interpretation of the free-surface
kinematic boundary condition is that a particle on the surface
always stays on the surface. The free-surface kinematic boundary
condition is a transport equation,  the free surface is advected,
or Lie transported (in the geometry literature), by the fluid
flow. Otherwise, the second equation in (\ref{sw2}) can be
interpreted as a continuity equation for the free surface, the
same as the continuity equation for the mass density in the
compressible fluids. With the formula of the Lie derivative of a
1-form along a vector field in view (see, for example, \cite{AbMa}
Section 2.2.),  the second equation in (\ref{sw2}) expresses the
fact that  the  1-form  $\mathrm{H}(x,t):=H(x,t)dx$ is advected,
or Lie transported, by the  vector field
$\mathrm{u}(x,t):=u(x,t)\pa_x$, that is, \be \frac{\pa
\mathrm{H}}{\pa t}+\mathrm{L}_{\mathrm{u}}\mathrm{H}=0,\label{lie}
\ee where $\mathrm{L}_{\mathrm{u}}$ denotes the Lie derivative
with respect to the vector field $\mathrm{u}$.

In the Lagrangian formalism one focuses the attention on the
motion of each individual particle of the mechanical system. We
denote by $M$ the ambient space whose points are supposed to
represent the particles at $t=0$. A diffeomorphism of $M$
represents the rearrangement of the particles with respect to
their initial positions. The set of all diffeomorphisms, denoted
Diff$(M)$, can be regarded (at least formally) as a Lie group. The
motion of the mechanical system is described by a time-dependent
family of orientation-preserving diffeomorphisms
$\gamma(\cdot,t)\in$ Diff($M$). For a particle
 initially located at $X$, the velocity at
 time $t$ is
\be
 \gamma_t(X,t):=\f{\pa \gamma(X,t)}{\pa t},\label{gamma}
 \ee this being the material velocity
 used in the Lagrangian description. The spatial velocity, used
 in the Eulerian description, is the flow velocity
\be
 u(x,t):=\gamma_t(X,t),\ee
$   \textrm{ at the location  } x=\gamma(X,t), \textrm{ at time }
t$, that is, \be
 u(\cdot,t)=\gamma_t\circ\gamma^{-1}.\label{u}
 \ee
 In the Lagrangian description, the velocity phase
 space is the tangent bundle $T\textrm{Diff}(M)$.
 In the Eulerian description,
 the spatial velocity is in the tangent space at the identity $Id$ of Diff$(M)$,
 that is, it is an element of the Lie algebra of Diff$(M)$.

In our problem,   $\mathrm{u}(x,t)$ can be regarded as a
time-dependent vector field on $\mathbf{R}$, that is, it belongs
to the Lie algebra of Diff$(\mathbf{R})$. Thus, in the Lagrangian
formalism of our problem we take $M=\mathbf{R}$
 and add the technical assumption that the
 smooth functions defined on  $\mathbf{R}$ with value in $\mathbf{R}$
  vanish rapidly at $\pm\infty$ together
 with as many derivatives as necessary. The configuration space of our problem
 is Diff$(\mathbf{R})$.  For a motion
 $\gamma(\cdot,t)\in$ Diff$(\mathbf{R})$ we have its Lagrangian velocity
 given by (\ref{gamma}) and its Eulerian velocity
 given by (\ref{u}).
 $\gamma$ is the flow of the
 time-dependent vector field $\mathrm{u}$.
 \\
The other unknown of our problem is $H(x,t)$, which for a fixed
$t$ can be regarded  as a real function on $\mathbf{R}$,
$H(\cdot,t)\in\mathcal{F}(\mathbf{R})$. We settle that the
evolution equation of $H(x,t)$ is the second equation in
(\ref{sw2}) which can be written  in the form (\ref{lie}). The
second equation in (\ref{sw2}) and the equation (\ref{lie}) are
equations in the Eulerian picture. With the aid of the pull back
map $\gamma^*$, we can write the Lagrangian form of the equation
(\ref{lie}), that is, \be \gamma^* \left(\frac{\pa \mathrm{H}}{\pa
t}+\mathrm{L}_{\mathrm{u}}\mathrm{H}\right)=0.\label{6}\ee We use
further  the interpretation of the Lie derivative of a
time-dependent 1-form along a time-dependent vector field in terms
of the flow of the vector field (see, for example, \cite{AbMa},
Section 2),  and we get the equation \be
\f{d}{dt}\left[\gamma^*(\mathrm{H})\right]=\gamma^*(\mathrm{L}_{\mathrm{u}}\mathrm{H})+
\gamma^*\left(\frac{\pa \mathrm{H}}{\pa
t}\right)\stackrel{(\ref{6})}{=} 0. \ee We denote this time
invariant 1-form in the reference configuration by \be
\mathrm{H}_0:=\gamma^*(\mathrm{H}),\quad
\mathrm{H}_0(X,t)=\mathrm{H}_0(X,0). \ee By the definiton of the
pull back map (see, for example, \cite{AbMa}, Section 2), we get
between the components of the 1-forms
$\mathrm{H}_0(X,t):=H_0(X,t)dX$ and $\mathrm{H}(x,t):=H(x,t)dx$
the following relation \be H_0= (H\circ \gamma)J_\gamma,\label{h0}
\ee where $J_\gamma:=\f{\pa \gamma}{\pa X}$ is the Jacobian of
$\gamma$, or, \be H= (H_0\circ
\gamma^{-1})J_{\gamma^{-1}}.\label{h} \ee

 In Lagrangian description,
 the equation of motion is the equation satisfied by a critical
 point of a certain functional $\mathfrak{a}(\gamma)$, called the action,
\be
 \mathfrak{a}(\gamma):=\int_0^T\mathcal{L}(\gamma,\gamma_t)dt,
 \ee
 defined
 on all paths  $\{\gamma(\cdot,t),$ $t\in[0,T]\}$ in $\textrm{Diff}(M)$,
  having fixed endpoints. $\mathcal{L}$ is a scalar function defined on
$T\textrm{Diff}(M)$, called Lagrangian.\\
 The Lagrangian for our
problem
   will be obtained by transporting the Lagrangian
   from the Eulerian picture, which by (\ref{energy}) and (\ref{21})
   is defined by 
\be \mathfrak{L}(u,H)=E_c(u)-E_p(H)=
\f1{2}\int_{-\infty}^\infty[u^2+u_x^2-(H-1)^2]dx,\label{lagreuler}
\ee
  to all  tangent spaces
  $T\textrm{Diff}(\mathbf{R})$, this transport being made taking
  into account
(\ref{u}) and (\ref{h}).\\
 For each function
$H_0\in\mathcal{F}(\mathbf{R})$ independent of time, we define the
Lagrangian
$\mathcal{L}_{H_0}:T\textrm{Diff}(\mathbf{R})\rightarrow\mathbf{R}$
by \be \mathcal{L}_{H_0}(\gamma,\gamma_t):=
\f1{2}\int_{-\infty}^\infty
\{(\gamma_t\circ\gamma^{-1})^2+[\partial_x(\gamma_t\circ\gamma^{-1})]^2-
[(H_0\circ \gamma^{-1})J_{\gamma^{-1}}-1]^2 \}dx.\label{lagr}\ee
 The Lagrangian $\mathcal{L}_{H_0}$ depends smoothly on $H_0$ and it is right  invariant
under the action of the subgroup \be
\textrm{Diff}(\mathbf{R})_{H_0}=\{\psi\in\textrm{Diff}(\mathbf{R})|
(H_0\circ \psi^{-1})J_{\psi^{-1}}=H_0\}
 \label{subgr}\ee that is, if we
replace the path $\gamma(t,\cdot)$ by
$\gamma(t,\cdot)\circ\psi(\cdot)$, for a fixed time-independent
$\psi$ in Diff($\mathbf{R}$)$_{H_0}$, then $\mathcal{L}_{H_0}$ is
unchanged.\\
The action on a path $\gamma(t,\cdot)$, $t\in [0,T]$, in
Diff($\mathbf{R}$) is
 \be
 \mathfrak{a}(\gamma):=\int_0^T\mathcal{L}_{H_0}(\gamma,\gamma_t)dt.\label{action''}\ee
The critical points of the action (\ref{action''}) in the space of
paths with fixed endpoints, satisfy \begin{equation}
\f{d}{d\varepsilon} \mathfrak{a}(\gamma+\varepsilon\varphi)\Big
|_{\varepsilon=0}=0,\label{critic}\end{equation} for every path
$\varphi(t,\cdot)$, $t\in[0,T]$, in $\textrm{Diff}(\mathbf{R})$
with endpoints at zero, that is,
$\varphi(0,\cdot)=0=\varphi(T,\cdot)$ and such that
$\gamma+\varepsilon\varphi$ is a small variation of $\gamma$ on
Diff($\mathbf{R}$). Taking into account (\ref{lagr}) and
(\ref{action''}), the condition (\ref{critic}) becomes
 \ba \hspace{-0.5cm}
\int_0^T\int_{-\infty}^{\infty}&&\left\{
\left(\gamma_t\circ\gamma^{-1}\right) \f{d}{d\varepsilon}\Big
|_{\varepsilon=0}\left[(\gamma_t+\varepsilon\varphi_t)\circ
(\gamma+\varepsilon\varphi)^{-1}\right
] \right.\nonumber\\
\hspace{-0.5cm} && +
\partial_x(\gamma_t\circ\gamma^{-1})\f{d}{d\varepsilon}\Big
|_{\varepsilon=0}\left[\partial_x\left((\gamma_t+\varepsilon\varphi_t)
\circ(\gamma+\varepsilon\varphi)^{-1}\right)\right]\nonumber\\
\hspace{-0.5cm} &&
-(H_0\circ\gamma^{-1})J^2_{\gamma^{-1}}\f{d}{d\varepsilon}\Big
|_{\varepsilon=0}[H_0\circ(\gamma+\varepsilon\varphi)^{-1}]\nonumber\\
\hspace{-0.5cm} &&
-(H_0\circ\gamma^{-1})^2J_{\gamma^{-1}}\f{d}{d\varepsilon}\Big
|_{\varepsilon=0}[J_{(\gamma+\varepsilon\varphi)^{-1}}]\nonumber\\
\hspace{-0.5cm} && +(J_{\gamma^{-1}})\f{d}{d\varepsilon}\Big
|_{\varepsilon=0}[H_0\circ(\gamma+\varepsilon\varphi)^{-1}]\nonumber\\
\hspace{-0.5cm} &&\left. +
(H_0\circ\gamma^{-1})\f{d}{d\varepsilon}\Big
|_{\varepsilon=0}[J_{(\gamma+\varepsilon\varphi)^{-1}}]
\right\}dxdt=0. \label{66'}\ea Let us now proceed to the
calculations of the terms in (\ref{66'}). For the   first two
terms in (\ref{66'}) see, for example, \cite{io}.\\
 Firstly,
differentiating with respect to $\varepsilon$ the identity \be
(\gamma+\varepsilon\varphi)\circ(\gamma+\varepsilon\varphi)^{-1}=Id
\ee one gets \be \f
d{d\varepsilon}\Big|_{\varepsilon=0}(\gamma+\varepsilon\varphi)^{-1}=
-\f{\varphi\circ\gamma^{-1}}{\gamma_x\circ\gamma^{-1}}.\label{7}\ee
We also have \be
\pa_x(\gamma_t\circ\gamma^{-1})=(\gamma_{tx}\circ\gamma^{-1})\pa_x(\gamma^{-1})=
\f{\gamma_{tx}\circ\gamma^{-1}}{\gamma_{x}\circ\gamma^{-1}},
\label{8}\ee \be
\pa_x(\varphi\circ\gamma^{-1})=(\varphi_{x}\circ\gamma^{-1})\pa_x(\gamma^{-1})=
\f{\varphi_{x}\circ\gamma^{-1}}{\gamma_{x}\circ\gamma^{-1}},
\label{9}\ee
 \ba
\pa_t(\varphi\circ\gamma^{-1})&=&\varphi_t\circ\gamma^{-1}+(\varphi_x\circ\gamma^{-1})
\pa_t(\gamma^{-1})\nonumber\\
&=&\varphi_t\circ\gamma^{-1}-
(\gamma_t\circ\gamma^{-1})\pa_x(\varphi\circ\gamma^{-1}),
\label{14}\ea \ba \hspace{-0.95cm}
\pa^2_x(\gamma_t\circ\gamma^{-1})&=&
\pa_x\left(\f{\gamma_{tx}\circ\gamma^{-1}}{\gamma_{x}\circ\gamma^{-1}}\right)
\nonumber\\
\hspace{-0.95cm}&=&\f{(\gamma_{txx}\circ\gamma^{-1})}{(\gamma_x\circ\gamma^{-1})^2}-
\f{(\gamma_{tx}\circ\gamma^{-1})(\gamma_{xx}\circ\gamma^{-1})}{(\gamma_x\circ
\gamma^{-1})^3},
 \label{10}\ea
\ba \hspace{-0.7cm} \f{d}{d\varepsilon}\Big
|_{\varepsilon=0}\left[\pa_x(\gamma+\varepsilon\varphi)^{-1}\right
]&=&\f{d}{d\varepsilon}\Big |_{\varepsilon=0}\left[\f
1{(\gamma_x+\varepsilon\varphi_x)\circ
(\gamma+\varepsilon\varphi)^{-1}}\right ]\nonumber\\
&=& - \f
{\pa_x(\varphi\circ\gamma^{-1})}{\gamma_x\circ\gamma^{-1}}+\f{\gamma_{xx}\circ\gamma^{-1}}{(\gamma_x\circ
\gamma^{-1})^3}(\varphi\circ\gamma^{-1}), \label{15} \ea \ba
\hspace{-0.7cm} \f{d}{d\varepsilon}\Big
|_{\varepsilon=0}\left[(\gamma_{tx}+\varepsilon\varphi_{tx})
\circ(\gamma+\varepsilon\varphi)^{-1}\right]&=&
\varphi_{tx}\circ\gamma^{-1}-(\gamma_{txx}\circ
\gamma^{-1})\f{\varphi\circ\gamma^{-1}}{\gamma_x\circ\gamma^{-1}}.
 \label{16}\ea
Taking into account (\ref{7}) - (\ref{16}) we obtain
 \ba
\f{d}{d\varepsilon}\Big
|_{\varepsilon=0}\left[(\gamma_t+\varepsilon\varphi_t)\circ
(\gamma+\varepsilon\varphi)^{-1}\right
]&=&\pa_t(\varphi\circ\gamma^{-1})+(\gamma_t\circ\gamma^{-1})\pa_x(\varphi\circ\gamma^{-1})\nonumber\\
&&-(\varphi\circ\gamma^{-1})\pa_x(\gamma_t\circ\gamma^{-1}),\label{18}
 \ea
 \begin{eqnarray}  \hspace{-1cm}\f{d}{d\varepsilon}\Big
|_{\varepsilon=0}\left[\partial_x\left((\gamma_t+\varepsilon\varphi_t)
\circ(\gamma+\varepsilon\varphi)^{-1}\right)\right] &=&
\pa_{tx}(\varphi\circ\gamma^{-1})+(\gamma_t\circ\gamma^{-1})\partial^2_x(\varphi\circ\gamma^{-1})\nonumber\\
\hspace{-1cm}&&-[\pa^2_x(\gamma_t\circ\gamma^{-1})](\varphi\circ\gamma^{-1}),
\label{12}\end{eqnarray} \ba \f{d}{d\varepsilon}\Big
|_{\varepsilon=0}[H_0\circ(\gamma+\varepsilon\varphi)^{-1}]&=&(H_{0_{x}}\circ\gamma^{-1})
\f{d}{d\varepsilon}\Big
|_{\varepsilon=0}(\gamma+\varepsilon\varphi)^{-1}\nonumber\\
&=&-(\varphi\circ\gamma^{-1})\pa_x(H_0\circ\gamma^{-1})
\label{17} \ea \ba \f{d}{d\varepsilon}\Big
|_{\varepsilon=0}[J_{(\gamma+\varepsilon\varphi)^{-1}}]&=&\f{d}{d\varepsilon}\Big
|_{\varepsilon=0}\left[\pa_x(\gamma+\varepsilon\varphi)^{-1}\right
]\nonumber\\
&=&- \f
{\pa_x(\varphi\circ\gamma^{-1})}{\gamma_x\circ\gamma^{-1}}+\f{\gamma_{xx}\circ\gamma^{-1}}{(\gamma_x\circ
\gamma^{-1})^3}(\varphi\circ\gamma^{-1})\nonumber\\
&=&-(J_{\gamma^{-1}})\pa_x(\varphi\circ\gamma^{-1})-\pa_x(J_{\gamma^{-1}})
(\varphi\circ\gamma^{-1}).\label{19} \ea
 Thus, with
(\ref{18})-(\ref{19}) in view,  the condition (\ref{66'})  becomes
\begin{eqnarray}
\int_0^T\int_{-\infty}^{\infty}&&
\left\{u\left[\partial_t(\varphi\circ\gamma^{-1})+u\partial_x(\varphi\circ\gamma^{-1})
-(\varphi\circ\gamma^{-1})u_x\right]\right.\nonumber\\
\hspace{1cm}&& + u_x\left[
\partial_{tx}(\varphi\circ\gamma^{-1})+u\partial^2_x(\varphi\circ\gamma^{-1})
-(\varphi\circ\gamma^{-1})u_{xx}\right]\nonumber \\
\hspace{1cm}&&
+HH_x(\varphi\circ\gamma^{-1})+H^2\pa_x(\varphi\circ\gamma^{-1})\nonumber\\
\hspace{1cm}&& -H_x
(\varphi\circ\gamma^{-1})-H\pa_x(\varphi\circ\gamma^{-1})\left.\right\}dxdt
=0,\label{20}\end{eqnarray} where $u=\gamma_t\circ\gamma^{-1}$ and
$H=(H_0\circ\gamma^{-1})J_{\gamma^{-1}}$.
 We integrate by parts with respect to
$t$ and $x$ in the above formula, we take into account that
 $u\rightarrow 0$,   $u_x\rightarrow 0$, $H\rightarrow 1$ at $\pm \infty$ and
$\varphi$ has endpoints at zero, and finally
 we get \begin{equation}
-\int_0^T\int_{-\infty}^\infty(\varphi\circ\gamma^{-1})\left[u_t+3uu_x
-u_{txx}- 2u_xu_{xx}-uu_{xxx}+HH_x\right]dxdt=0 \end{equation}
Therefore, to the order of our approximation, we proved:

\vspace{0.3cm}

\textbf{Theorem 1.} \textit{For an irrotational
  shallow water flow,   the non-dimensional horizontal velocity
 of the water $u(x,t)$ and the non-dimensional free upper surface
$H(x,t)=1+\epsilon\eta(x,t)$,
  satisfy the Camassa-Holm shallow water system (\ref{2ch}) with the  plus sign}.

\vspace{0.3cm}

\textbf{Remark}. We observe that, if instead of the Lagrangian
(\ref{lagreuler}) we consider the $H^1$ metric in $u$ and $L^2$
metric in $(H-1)$, that is, \be
\f1{2}\int_{-\infty}^\infty[u^2+u_x^2+(H-1)^2]dx, \ee by
transporting this metric to all tangent spaces
  $T\textrm{Diff}(\mathbf{R})$, the transport being made taking
  into account
(\ref{u}) and (\ref{h}), we  get,  for each function
$H_0\in\mathcal{F}(\mathbf{R})$ independent of time,
$\mathcal{E}_{H_0}:T\textrm{Diff}(\mathbf{R})\rightarrow\mathbf{R}$
defined by  \be \mathcal{E}_{H_0}(\gamma,\gamma_t):=
\f1{2}\int_{-\infty}^\infty
\{(\gamma_t\circ\gamma^{-1})^2+[\partial_x(\gamma_t\circ\gamma^{-1})]^2+
[(H_0\circ \gamma^{-1})J_{\gamma^{-1}}-1]^2 \}dx\label{lagr'} \ee
$\mathcal{E}_{H_0}$ depends smoothly on $H_0$ and its right
invariant under the action of the subgroup  (\ref{subgr}). The
critical points of the action \be
\int_0^T\mathcal{E}_{H_0}(\gamma,\gamma_t)dt, \ee in the space of
paths with fixed endpoints,  will satisfy
\begin{equation}
-\int_0^T\int_{-\infty}^\infty(\varphi\circ\gamma^{-1})\left[u_t+3uu_x
-u_{txx}- 2u_xu_{xx}-uu_{xxx}-HH_x\right]dxdt=0 \end{equation}
Thus, the functions $u(x,t)$ and  $H(x,t)$  will fulfill
   the Camassa-Holm  system (\ref{2ch}) with the sign minus.
We point out that (\ref{2ch}) with the minus sign can be regarded
as geodesic equations on the semidirect product of
Diff$(\mathbf{R})$ with $\mathcal{F}(\mathbf{R}$) (see also
\cite{honatr}).







\end{document}